\documentclass{aa}

\begin{document}

   \thesaurus{     
              12.12.1;  %
             13.07.1} %
   \title{An intrinsic anisotropy in the angular distribution of gamma-ray
          bursts}

   \author{L.G. Bal\'azs    \inst{1}
     \and  A. M\'esz\'aros  \inst{1,2,3}
     \and  I. Horv\'ath     \inst{4}
     \and  R. Vavrek        \inst{1} }

   \offprints{L.G. Bal\'azs}

   \institute{Konkoly Observatory, Budapest, Box. 67, H-1505 Hungary\
              (balazs@ogyalla.konkoly.hu, vavrek@ogyalla.konkoly.hu)
              \and Department of Astronomy, Charles University,
              V Hole\v{s}ovi\v{c}k\'ach 2, CZ-180 00 Prague 8,
              Czech Republic\\
              (meszaros@mbox.cesnet.cz)
              \and European Southern Observatory, Karl-Schwarzschild-Strasse 2,
              Garching bei M\"unchen, Germany\
              \and Department of Physics, BJKMF,
              Budapest POB.12, H-1456 Hungary\
              (hoi@rmki.kfki.hu) }

   \date{Received <date> / Accepted <date>}

   \maketitle

   \begin{abstract}

   The anisotropy of the sky distribution of 2025 gamma-ray bursts (GRBs)
   collected in Current BATSE catalog is confirmed. It is shown that
   the quadrupole term being proportional to $\sim \sin 2b \sin l$ is
   non-zero with a probability 99.9\%. The occurrence of this ani\-sot\-ropy
   term is then supported by the binomial test even with the probability
   99.97 \%. It is also argued that this anisotropy cannot be caused
   exclusively by instrumental effects due to the non-uniform sky exposure
   of BATSE instrument; there should exist also some intrinsic anisotropy
   in the angular distribution of GRBs. Separating GRBs into short
   and long subclasses, it is shown that the 251 short ones are distributed
   anisotropically, but the 681 long ones seem to be distributed still
   isotropically. The 2-sample Kolmogorov-Smirnov test shows that they are
   distributed differently with a 98.7\% probability.
   The character of anisotropy suggests that the cosmological
   origin of short GRBs further holds, and there is no evidence for their
   Galactical origin. The work in essence contains the key ideas and
   results of a recently published paper (\cite{balazs}),
   to which the new result following from
   the 2-sample Kolmogorov-Smirnov test is added, too.

   \end{abstract}

   \section{Mathematical considerations}

   From the mathematical point of view, the necessary condition for
   the isotropy is the stochastic independency of the sky distribution
   of the bursts on their observed physical properties. This means that,
   if $f(b,l, x_1,...,x_n)$ $dF dx_1...dx_2$ is the probability of finding
   an object in the $dF=\cos b\; dl\;db$ infinitesimal solid angle and
   in the $(x_1, x_1+dx_1, ..., x_n,x_n+dx_n)$ interval, one must have
   \begin{equation}
   f(l,b,x_1,...,x_n) = \omega(l,b) g(x_1,...,x_n).
   \end{equation}
   Here $0\leq l \leq 360^o,\; -90^o \leq b \leq 90^o$ give the celestial
   positions in Galactical coordinates, $x_n$ ($n \geq 1$) measure the
   physical properties (peak fluxes, fluences, durations, etc...) of GRBs
   and $g$ is their probability density. One may assume the fulfilment of
   this equation for GRBs.

   In the case of spatial isotropy, assuming that the detection
   probability does not depend on the celestial direction, one has:
   $\omega(l,b)=1/(4\pi) $. In general case one may decompose the
   function $\omega(b,l)$ into the spherical harmonics. One obtains:
   $$
   \omega(b,l) = (4\pi)^{-1/2} \omega_0 - $$
   $$ (3/(4\pi))^{1/2} (\omega_{1,-1} \cos b \sin l
   - \omega_{1,1} \cos b \cos l
   +  \omega_{1,0} \sin b) +$$
   $$(15/(16\pi))^{1/2}( \omega_{2,-2} \cos^2 b \sin 2l
   +  \omega_{2,2} \cos^2 b \cos 2l - $$
   $$ \omega_{2,-1} \sin 2b \sin l
   -   \omega_{2,1} \sin 2b \cos l) + $$
   \begin{equation}
   (5/(16\pi))^{1/2}\omega_{2,0} (3 \sin^2 b - 1)  +  higher\; order\;
   harm.
   \end{equation}

   The first term on the right-hand side is the monopole term, the
   following three ones are the dipole terms, the following five ones
   are the quadrupole terms. Since $\omega$ is constant for isotropic
   distribution, on the right-hand side any terms, except for $\omega_0$,
   should be identically zeros. To test this hypothesis one
   has to compute
   the values of the corresponding spherical harmonic at the celestial
   positions of the observed GRBs and apply,
   e.g., the Student test in order to see that the mean of the computed
   values significantly differs from zero.

   A further trivial consequence of the isotropy is the expected equal
   number of bursts in celestial regions of equal areas. For example, one
   may divide the celestial sphere into two equal areas, e.g., taking
   those regions in which the sign of a given harmonic is either positive
   or negative, respectively.
    Then one may compare the number of GRBs in these regions
   by the standard binomial (Bernoulli) test. The details
   (together with the relevant references) of this test
   and also of the test based on the spherical harmonics are discussed
   in \cite{balazs}.

   \section{On the existence of the intrinsic anisotropy}

   In order to test the isotropy of 2025 GRBs we test the three dipole
   and five quadrupole terms. One obtains that, except for the terms
   defined by $\omega_{2,-1}$ and $\omega_{2,-2}$, the remaining six
   terms may still be taken to be zero. This means that there is a clear
   anisotropy defined by term $\sim \sin 2b \sin l$. The probability that this
   term is zero is $0.1\%$. In addition, the second quadrupole
   term being proportional to $\cos^2b \sin 2l$ is non-zero, too, with the
   probability $0.6\%$.

   A straightforward counting of GRBs in those regions of equal areas, where
   $\sin 2b \sin l$
   has either positive or negative signs, respectively, shows that 930
   GRBs are in the first area and 1095 are in the second one. Taking
   $p=0.5$ probability for the binomial (Bernoulli) test, one obtains a
   $0.03\%$ probability that this detected distribution is caused by
   a chance. The observed distribution of all GRBs on sky is anisotropic
   with a certainty.

   Instrumental effects of BATSE instrument should also
   play a role, since the sky exposure of BATSE instrument is non-uniform.
   The known dependence of the detection probability of this instrument
   predicts a similar kind of anisotropy. Hence,
   the question is the following: Is this anisotropy caused either
   exclusively by the non-uniform sky-exposure function of BATSE
   instrument, or is there also an intrinsic anisotropy in the
   distribution of GRBs?

   To clarify the situation we divided the GRBs into two groups
   according to their durations $T_{90}$. We
   excluded the dimmest GRBs. Then, 932 GRBs were
   separeted into the "short" ones (251 GRBs; $T_{90}<2$ s), and
    "long" ones (681 GRBs; $T_{90}>2$ s).

   Dividing the sky into the two equal areas, as described above, we
   obtain a different behaviour for the short and long GRBs, respectively
   (see Table 1.).

   \begin{table}
    \caption{Results of the binomial test of subsamples of GRBs
     with different durations. $N$ is the number of GRBs at the
     given subsample, $ k_{obs}$ is the observed number GRBs at
     the first area for this subsample, and \% is the probability
     in percentages that the assumption of isotropy still holds.}
     \label{Short-long}
   $$
         \begin{array}{ccrcc}
            \hline
            \noalign{\smallskip}
            sample & N & k_{obs} & (N-k_{obs}) & \% \\
            \noalign{\smallskip}
            \hline
            \noalign{\smallskip}
   all \phantom{@}GRBs & 932 & 430 & 502 &  2.0 \\
   T_{90}<2s & 251  & 103 & 148  &  0.55  \\
   T_{90}>2s & 681  & 327 & 354 & 32 \phantom{@}  \\
   \hline
   \end{array}
   $$
   \end{table}

   Table 1. shows that the short GRBs are further distributed
   anisotropically; there is a smaller than $1\%$ probability of isotropy.
   On the other hand, the long GRBs can still be distributed
   isotropically. Up to this point all results mentioned in this Section,
   were also written down in \cite{balazs} together with the relevant
   references.

   The application of the 2-sample Kolmogorov-Smirnov test (\cite{press})
   on  $\omega_{2,-1}$ shows that the significance of the difference among
   the samples of short and long GRBs is $98.7\%$. The short and long ones
   are obviously distributed differently with a probability $98.7\%$.
   Note here that this important
   quantification of the different behaviour of two subclasses is a new result
   not presented in \cite{balazs}.

   We mean that these values confirm the expectation that there must exist
   some intrinsic anisotropy in the distribution of GRBs. Once there
   were an exclusive instrumental origin of the anisotropy all GRBs, the
   character of anisotropy should be the same for both types of GRBs;
   there should exist no difference among the short and long samples.
   Of course, the character of anisotropy is quite different than expected
   for the Galactical origin. For this one would need a clear non-zero
   $\omega_{2,0}$ spherical harmonic. Hence, there is no doubt concerning
   the cosmological origin of GRBs.

   \section{Conclusions}

   \begin{enumerate}
      \item  The distribution of 2025 GRBs is anisotropic with a
             99.97\% probability.

      \item Separating GRBs into the two classes it is shown that
            the short ones are distributed anisotropically with a
            higher than 99\% probability, but the long ones can still
            be distributed isotropically. There is a 98.7\% probability
            that the short and long GRBs are distributed differently.

      \item This different behaviour of two samples cannot be caused
            exclusively by instrumental effects; there must exist also
            some intrinsic anisotropy, too.

      \item The cosmological origin is not queried by this ani\-sot\-ro\-py.
   \end{enumerate}

   \begin{acknowledgements}
        A.M. thanks for the kind hospitality at ESO.
       R.V. acknowledges the valuable discussions with I. Domsa.
       This article was partly supported by OTKA grants T024027
(L.G.B.), F024027 (I.H.), by GAUK grant 36/97 and by GA\v{C}R grant
202/98/0522 (A.M.).
   \end{acknowledgements}

\end{document}